# Mössbauer studies on FeSe and FeTe

Yoshikazu Mizuguchi[a,b,c], Takao Furubayashi[a], Keita Deguchi[a,b,c], Shunsuke Tsuda[a,b], Takahide Yamaguchi[a,b] and Yoshihiko Takano[a,b,c]

[a]Superconducting Materials Center, National Institute for Materials Science, Tsukuba, 305-0047, Japan
[b]JST, Transformative Research-Project on Iron Pnictides, Tsukuba, 305-0047, Japan
[c]Graduate School of Pure and Applied Sciences, University of Tsukuba, Tsukuba, 305-0006, Japan

Corresponding author: Yoshikazu MIZUGUCHI
E-mail address: MIZUGUCHI.Yoshikazu@nims.go.jp
Address: 1-2-1 Sengen, Tsukuba 305-0047, Japan
Phone: +81-29-859-2842
Fax: +81-29-859-2601

Abstract
We carried out $^{57}$Fe Mössbauer measurements for FeSe and $Fe_{1.08}Te$ to investigate the magnetic properties. There was no sign of magnetic ordering above 4.2 K for superconducting FeSe. The magnetic sextet corresponding to antiferromagnetic ordering of Fe in low-spin state was observed for non-superconducting $Fe_{1.08}Te$.

1. Introduction

Since the discovery of LaFeAsO$_{1-x}$F$_x$ superconductor [1], various iron-based superconductors, which have a layered structure analogous to LaFeAsO, have been discovered. Iron-based superconductivity has attracted researchers all over the world due to its high transition temperature $T_c$ while they are mainly composed of iron which has been considered to be disadvantageous for superconductivity because it is magnetic. In this respect, iron-based superconductivity attracts us by not only high $T_c$ but also the mechanism of superconductivity.

Among the iron-based superconductors, FeSe has the simplest crystal structure [2]. It shows dramatic enhancement of $T_c$ under high pressure. The $T_c$ increases from 12 to 37 K [3-5]. NMR measurements indicated an enhancement of antiferromagnetic fluctuations under high pressure where the $T_c$ was dramatically enhanced [6,7]. While FeSe shows high $T_c$, an analogue compound FeTe does not show superconductivity but undergoes antiferromagnetic ordering at 70 K [8–10]. To investigate the magnetic states of FeSe and FeTe, we carried out $^{57}$Fe Mössbauer measurements.

2. Materials and methods

Polycrystalline samples of FeSe, Fe$_{1.08}$Te were synthesized using a solid-state reaction method as described in ref. 3 and 10. The FeSe sample contains a minor phase of hexagonal FeSe, which undergoes magnetic ordering [11]. We carried out the $^{57}$Fe mössbauer measurements for FeSe and FeTe$_{1.08}$ from room temperature to 4.2 K.

3. Results and discussions

Figure 1 shows the $^{57}$Fe Mössbauer spectra of FeSe at 300, 150, 77 and 4.2 K. Major feature of the spectra can be explained by a single paramagnetic doublet. For the major phase of superconducting FeSe, the spectrum obtained at 4.2 K shows no sign of magnetic ordering at 4.2 K. In fact, FeSe does not exhibit magnetic ordering down to 4.2 K while it undergoes a structural transition from tetragonal to orthorhombic around 70 K [12]. An Isomer shift (*IS*) and a quadrupole splitting (*QS*) are estimated to be 0.5380(60) mm/s and 0.268(10) mm/s, respectively. These values are almost consistent with the previous report [13]. The spectra contain the signals attributed to the impurity phases of hexagonal FeSe and tiny amounts of iron oxides. The small magnetic sextets, which is fitted with hyperfine field $H_f$ = 264.7(80) kOe, will corresponds to the signal of hexagonal FeSe.

Figure 2 shows the $^{57}$Fe Mössbauer spectra of Fe$_{1.08}$Te at room temperature, 77 and 4.2 K. The spectrum at room temperature is analyzed by two paramagnetic doublets.

The main doublet, which has the parameters of $IS$ = 0.452(12) mm/s and $QS$ = 0.315(28) mm/s, will correspond to the paramagnetic Fe of the FeTe layers. The estimated $IS$ is the value close to that of FeSe. The minority could be attributed to the excess Fe which exists at the interlayer site. For the spectrum at 4.2 K, a clear hyperfine magnetic sextet is observed, corresponding to the magnetic ordering below 70 K. The $H_f$ is estimated to be 103.4(11) kOe. Considering the comparably low field, the state of the Fe spin should be low-spin state.

To summarize, we carried out the $^{57}$Fe Mössbauer measurements for FeSe and Fe$_{1.08}$Te. The single paramagnetic doublet was observed at whole temperatures for the superconducting FeSe phase. For Fe$_{1.08}$Te, the clear magnetic sextet was observed at 4.2 K, which was consistent with the antiferromagnetic ordering of Fe in the low-spin state associated with the structural change at 70 K.

Acknowledgements

This work was partly supported by a Grand-in-Aid for Scientific Research (KAKENHI).

Fig. 1

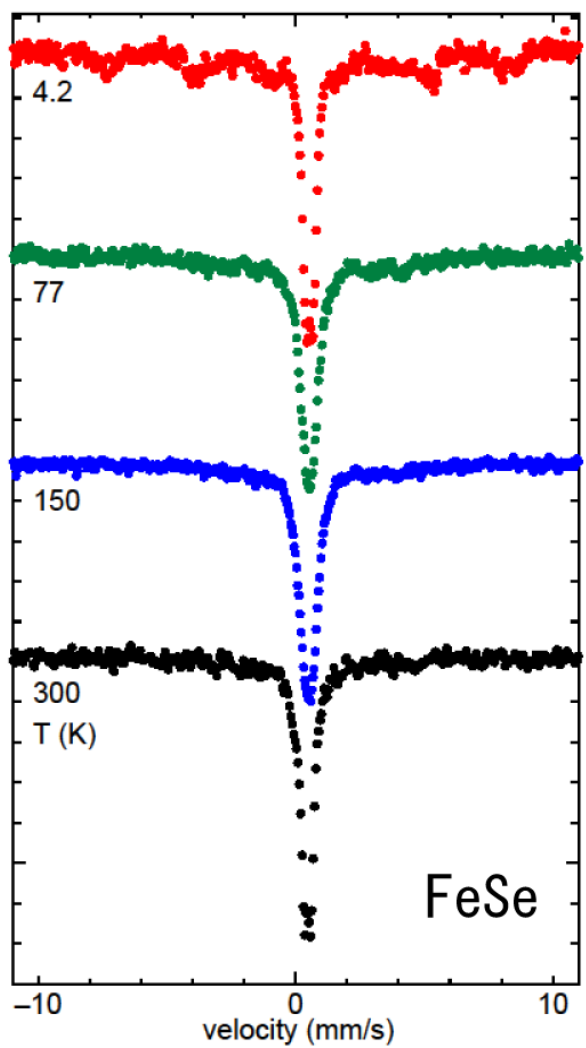

Fig. 1. $^{57}$Fe Mössbauer spectra of FeSe at 300, 150, 77 and 4.2 K. There is no sign of magnetic ordering for the major phase of tetragonal FeSe.

Fig. 2

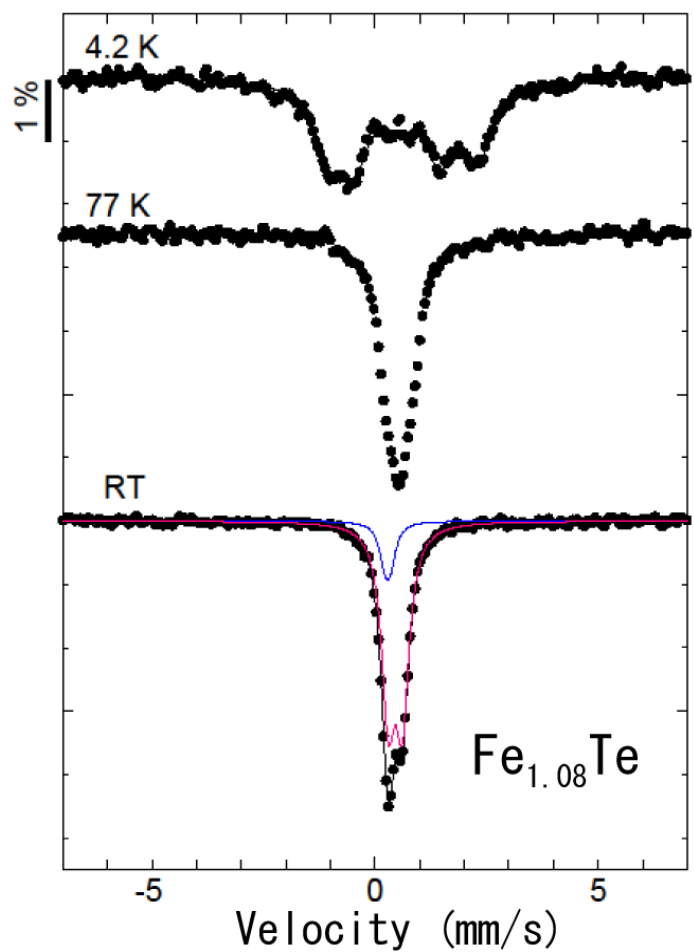

Fig. 2. $^{57}$Fe Mössbauer spectra of Fe$_{1.08}$Te at room temperature, 77 and 4.2 K. The spectra are fitted using two types of doublets, which will be attributed to the two Fe sites. One is the Fe of the FeTe layer, and the other is Fe which exists at the interlayer site.